\documentclass[prl,amsmath,floatfix,reprint]{revtex4-1}
\usepackage{graphicx}
\newcommand\eg{\textit{e.g.}}
\renewcommand\Re{\mathrm{Re}}
\graphicspath{{../}}

\begin{document}
\title[Contrasting physics of laminar and turbulent flow]{An elementary example of contrasting laminar and turbulent flow physics}
\author
 {Paolo Luchini}
 \email{luchini@unisa.it}
\affiliation
{DIIN, University of Salerno, Italy}
\begin{abstract}
Something as simple as Couette and Poiseuille onedimensional flow of a newtonian fluid between infinite parallel walls provides an illuminating example of the contrasting physics of laminar and turbulent flow: the difference between their mean velocity profiles has in one regime the opposite sign than in the other. This easily verifiable but yet unnoticed behaviour implies that the physical mechanisms of laminar and turbulent transport are even more fundamentally different than ordinarily presumed.
\end{abstract}
\maketitle

\section{Introduction}
Nobody in the physics of turbulence community would dispute that the transport of momentum, giving rise to the dissipation of mechanical energy in turbulent flow of a fluid around or inside solid bodies, has its own distinguishing features that set it apart from classical momentum diffusion in laminar viscous flow: laminar diffusion is caused by local, small-scale interactions at the molecular level whereas an essential feature of turbulence is its wide range of significant length and time scales. Nonetheless one would hardly expect the qualitative difference between turbulent and laminar transport to be as large as to reverse the effect of a pressure gradient upon the flow.

In a recent sequence of papers\cite{PRL, APS, structure, iTi}, we have been investigating the universality of the classical logarithmic law of the mean turbulent velocity profile, and the experimental discrepancies from this law that did not fail to arouse sometimes spirited controversies in past and recent times. In fact, to discriminate classical versus alternate theories of the mean turbulent velocity profile is the goal of still ongoing dedicated large-scale experiments \cite{superpipe,ciclope,HiReff}. In this context we found that a representative collection of numerical and experimental data can be accurately fitted in the classical wall-layer and defect-layer matched asymptotics, or equivalently represented as the sum of a ``law of the wall'' and a ``law of the wake'' in the language of Coles \cite{Coles}, provided a) the law of the wake is allowed to vary from one geometry to another and b) the portion of the law of the wake superposing with the logarithmic layer is linear in the wall-normal coordinate z, just as dimensional analysis dictates when it is assumed to be linear in the pressure gradient \cite{PRL}. The interpolations provided \cite{structure} of both the law of the wall and the law of the wake satisfactorily predict the available data, and particularly well predict the recent Hi-Reff pipe-flow experiment \cite{HiReff} which had not been used in the process of fitting.

Purpose of the present letter is to expose an extremely simple but not intuitive property of turbulent plane Couette and Poiseuille flow which naturally springs out as a corollary of the above analysis, and favourably synergizes with some other previous research of ours about the effects of external forcing upon a turbulent flow.

\section{The relative height of Couette and Poiseuille velocity profiles}
\begin{figure*}
\begin{tabular}{c@{}c}
laminar & turbulent\\
\includegraphics[width=0.5\textwidth]{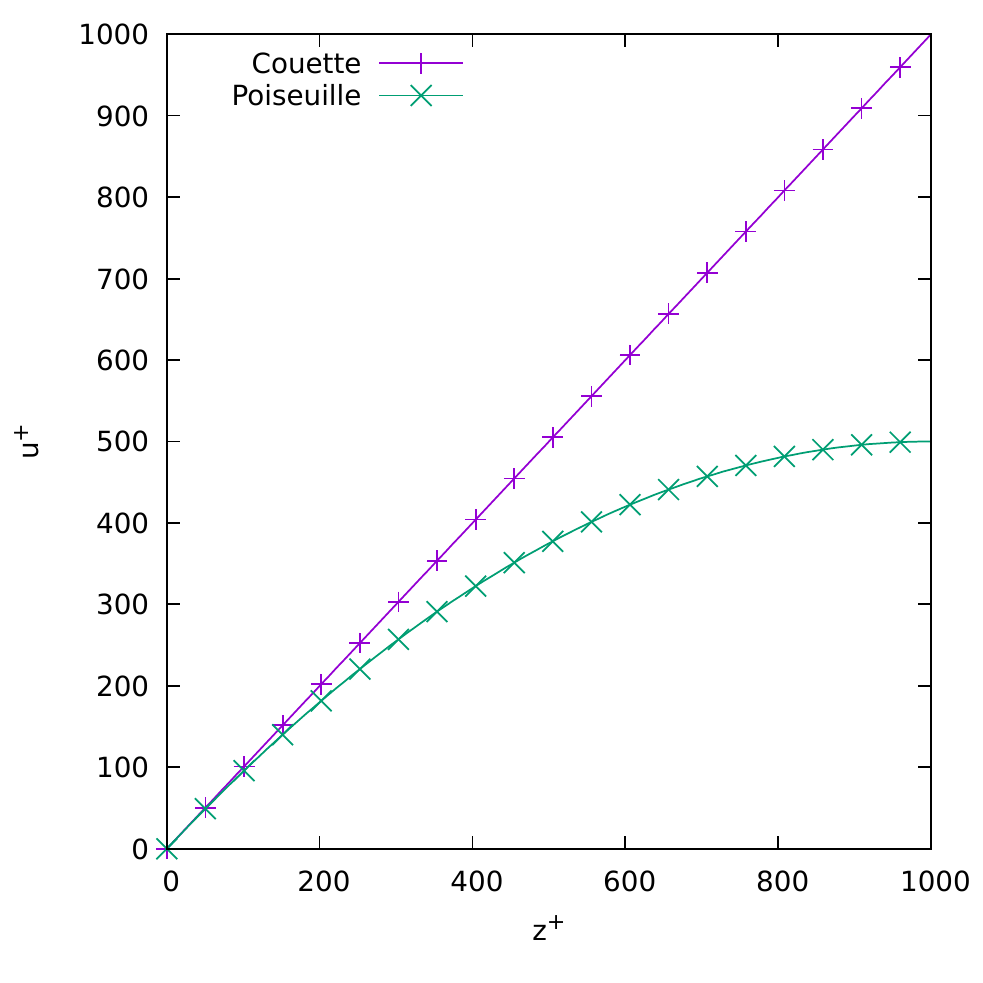}&
\includegraphics[width=0.5\textwidth]{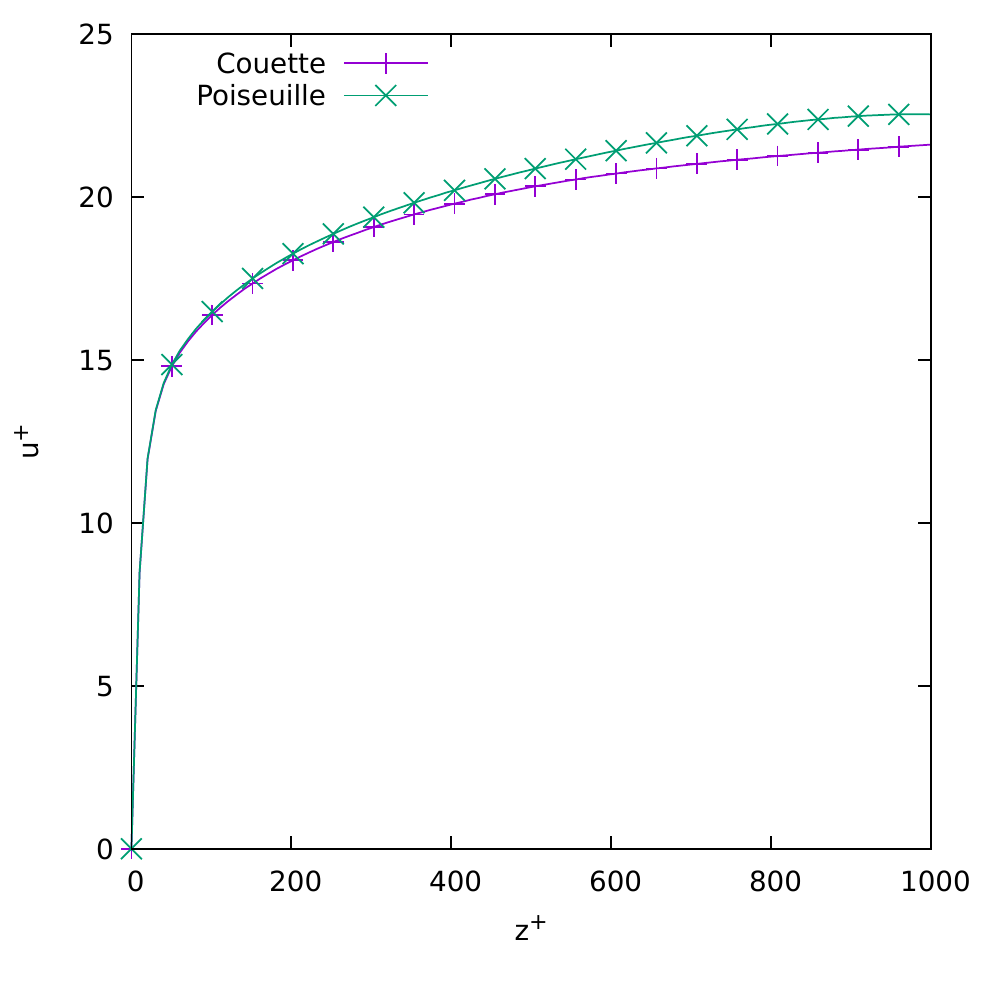}\\
\end{tabular}
\caption{Couette and Poiseuille velocity profiles are oppositely stacked in laminar and turbulent flow.}
\label{pgcomp}
\end{figure*}

The observed behaviour of the mean velocity profile in different geometries of flow at the same Reynolds number was exemplified in Figure 1 of \cite{PRL}, based on the numerical data of \cite{Pirozzoli} and \cite{Moser}. It clearly appeared from this figure, and can be seen again in the present Figure \ref{pgcomp} (right), that the velocity profile of turbulent plane Poiseuille flow, in wall units, sits consistently above the velocity profile of plane Couette flow all the way up to the centerline. Whereas these plots were drawn at a single Reynolds number $\Re_\tau=1000$, as data for Couette flow were not available at larger Reynolds numbers, the theory set out in \cite{PRL, structure} makes it plausible for this to be a generic behaviour to be found again at any other Reynolds number. In fact if each velocity profile is decomposed into a law of the wall and a law of the wake, the law of the wall being the same, the difference of the two velocity profiles is nothing else than the difference of the two laws of the wake, independent of the Reynolds number in outer units. In the light of the success of this decomposition in all the situations investigated in \cite{structure}, we can assert it as a fact that the turbulent incarnation of plane Poiseuille flow always has a \emph{higher} mean velocity than turbulent Couette flow in wall units. An equivalent statement in physical units is that Poiseuille flow exhibits a higher velocity than Couette flow when compared for the same value of shear stress at the wall.

The relative ordering of turbulent Couette and Poiseuille velocity profiles might look unsurprising, but it becomes a striking observation if one notices that it is opposite to the ordering of the corresponding laminar flows, where for the same wall-shear stress the velocity profile of Poiseuille flow is everywhere \emph{lower} than in Couette flow.
The laminar outcome is fairly obvious in its own right:
if both velocity $u$ and wall-normal coordinate $z$ are normalized in wall units \footnote{Thus becoming $u^+=u/u_\tau$ and $z^+=z u_\tau/\nu$ with $u_\tau=\sqrt{\tau_w/\rho}$, $\tau_w$ being the wall shear stress, $\rho$ the density and $\nu$ the kinematic viscosity. The Reynolds number is then $\Re_\tau=h u_\tau/\nu$, $h$ being the half-height of the channel.}, the derivative of the velocity profile at the wall is constrained to be unity; the addition of a favourable pressure gradient induces a downwards curvature and lowers the laminar Poiseuille velocity profile below the Couette velocity profile, in the manner represented in Figure \ref{pgcomp} (left) for the particular Reynolds number $\Re_\tau=1000$.
At the same Reynolds number, turbulent Couette and Poiseuille flow rank instead as in Figure \ref{pgcomp} (right), with a difference which is smaller but opposite in sign. The same will also be true at any other Reynolds number, if confidence is accorded to our previous results.

\section{Can turbulent transport be modelled as inhomogeneous diffusion?}
As recalled in the introduction, every involved scientist would agree that laminar and turbulent transport are qualitatively different. Nonetheless, where quantitative modelling of the mean flow is needed some form of ``eddy viscosity'' is in practice always invoked. Eddy viscosity reduces the action of turbulence to an effective space-varying classical viscosity, based on the idea that turbulent eddies randomly redistribute momentum with spatially changing intensity, and ubiquitously appears in both numerical codes and theoretical approximations. In mechanics and aeronautics eddy viscosity is usually a function of a set of auxiliary dynamical variables, typicaly turbulent energy and dissipation or an extension thereof \citep[\eg][]{Wilcox}. Simpler mixing-length models where eddy viscosity is a given function of distance to the wall, or for the purpose of gross simplification even sometimes a constant, are pervasive in geophysical applications such as atmospheric and marine flow over topography \citep[including theoretical analyses, \eg][]{BH}.

A paradox arising from the adoption of eddy viscosity was pointed out when my student Serena Russo and I \cite{TesiRusso, RL} undertook a numerical investigation of flow over slowly undulating topography modelled through an equivalent volume force, akin to the one giving rise to steady streaming \cite{streaming}. To our own surprise, we found out that a volume force $F(z)$ with zero integral, or resultant, such as not to modify the wall shear stress for a given pressure gradient, produces a flow rate of opposite direction in laminar and in turbulent flow. Since the flowrate perturbation predicted by any (positive) eddy viscosity can be proved to be congruent with the laminar result, it turns out that the prediction is of the wrong sign, and remains so for any possible eddy-viscosity model.
 The same conclusion was later confirmed by a direct numerical simulation fully including the sinusoidal wall in the form of an immersed boundary \cite{imb}, and indirectly by a comparison with much earlier experiments \cite{Hanratty,addendum}.


All of a sudden a connection appears: the present Figure \ref{pgcomp} highlights an opposite reaction of laminar and turbulent flow to a pressure gradient, or equivalently to a constant force, just as Figures 2 and 11 of \cite{RL} showed an opposite reaction of laminar (or eddy-viscosity) and turbulent flow to a force profile with zero mean. It thus turns out that the reversal between Poiseuille and Couette profiles depicted in Figure \ref{pgcomp} is an even simpler piece of evidence of the same counterintuitive phenomenon studied in \cite{RL}, and can be construed as an even more fundamental example of the inadequacy of modelling turbulent transport as a form of inhomogeneous laminar diffusion.

Specifically the turbulent Couette profile $u_C(z)$, or at least some portion of it nearest to the wall, can be continuously morphed into the turbulent Poiseuille velocity profile $u_P(z)$ by gradually adding a pressure gradient. It is consistent with the usual eddy-viscosity approximations to describe the perturbation $\delta u=u_P(z)-u_C(z)$ by a linearized model, much as is standard practice in dealing with boundary perturbations to a parallel flow \cite{BH}, where $\delta u$ obeys the Reynolds-averaged equation $p_x=\tau_z$ with $\tau/\rho=\nu_T(z)\delta u_z$ and a constant (negative, favourable) $p_x$, or equivalently
\begin{equation}
\left[\nu_T(z)\delta u_z\right]_z = p_x/\rho \tag{1}.
\label{eq1}
\end{equation}
Here $\nu_T(z)$ is the eddy viscosity, reducing to molecular viscosity for $z=0$ and elsewhere a function of properties of the flow (the velocity gradient, turbulent energy, possibly others) which however numerous must eventually become given functions of $z$ in a $z$-only dependent unperturbed flow. For a wall shear stress imposed to be the same, the normal derivatives of $u_P(z)$ and $u_C(z)$ at the wall coincide; then the boundary conditions for (\ref{eq1}) are $\delta u(0)=\delta u_z(0)=0$.

The solution $\delta u$ of (\ref{eq1}) for a favourable, negative pressure gradient and any positive eddy viscosity is bound to be negative, just as it is when viscosity is constant in laminar flow, whereas the true turbulent $\delta u$ displayed in Figure \ref{pgcomp} is positive. Therefore the conclusion attained by Russo and Luchini \cite{RL} with reference to a zero-mean volume force applies just as well to the mean pressure gradient:
the widespread eddy-viscosity model \eqref{eq1} is an oversimplification and fails to capture the observed onedimensional physics.

\section{Conclusion}
Purpose of this letter was to demonstrate an astoundingly simple physical example of the contrasting behaviour of laminar and turbulent flow, an example which stems from the convergence of two previous results of ours and makes them reinforce each other: the difference between the mean velocity profiles of plane Poiseuille and Couette flow, for the same wall shear stress, has in the turbulent regime the opposite sign than in the laminar regime. In physical terms the addition of a negative pressure gradient, and the wall-normal shear-stress decrease that it induces according to $p_x=\tau_z$, affects laminar and turbulent flow in opposite directions. A local proportionality between velocity gradient and shear stress cannot account for this: some degree of nonlocality is something that all future quantitative descriptions of turbulent shear flow will have to account for. Many past ones do not.

\end{document}